\documentclass[aps,pre,twocolumn]{revtex4-1}

\usepackage{amsmath}

\usepackage{color}

\usepackage{graphicx}

\DeclareMathAlphabet{\mathitbf}{OML}{cmm}{b}{it}

\newcommand{\ket}[1]{|#1\rangle}
\newcommand{\bra}[1]{\langle #1|}
\newcommand{\braket}[2]{\langle #1|#2\rangle}

\begin{document}
\title{Breakdown of continuum elasticity in amorphous solids}
\author{ Edan Lerner${}^1$, Eric DeGiuli${}^1$, Gustavo D\"uring${}^2$, and Matthieu Wyart${}^1$}
\affiliation{${}^1$ New York University, Center for Soft Matter Research, 4 Washington Place, New York, NY, 10003, USA \\
${}^2$ Facultad de F\'isica, Pontificia Universidad Cat\'olica de Chile, Casilla 306, Santiago 22, Chile}

\begin{abstract}
We show numerically that the response of simple amorphous solids (elastic networks and particle packings) to a local force dipole is characterized by a lengthscale $\ell_c$ that diverges as unjamming is approached as $\ell_c \sim (z - 2d)^{-1/2}$, where $z \ge 2d$ is the mean coordination, and $d$ is the spatial dimension, at odds with  previous numerical claims. We also show how the magnitude of the lengthscale $\ell_c$ is amplified by the presence of internal stresses in the disordered solid. Our data suggests a divergence of $\ell_c\sim (p_c-p)^{-1/4}$ with proximity to a critical internal stress $p_c$ at which soft elastic modes become unstable. 
\end{abstract}

\maketitle

\section{introduction}

At long wavelength, amorphous solids behave as isotropic elastic solids. At short wavelength,  however, this continuum description breaks down, and the particle-scale disorder matters. This fact is well-known in granular materials where the response to a local perturbation leads to a heterogeneous response locally, and where the stress propagates along preferred paths, or force chains \cite{majmudar,majumdar2}. In molecular glasses, the breakdown of a hydrodynamic description is visible in the density of vibrational modes, which departs from the Debye prediction (valid in the continuum) at frequencies typically about a tenth of the Debye frequency. At such frequencies the density of vibrational modes is larger than expected, a phenomenon referred to as the boson peak  \cite{phillips_book}. Converting the boson peak frequency to a length scale using the transverse speed of sound leads to a length scale on the order of ten particle diameters \cite{sette,mossa}. What governs this length scale is debated  \cite{lubchenkoorigin,schirmacher,Wyart05}.% Our aim is to clarify this question in simple amorphous solids. 

Understanding amorphous solids at such intermediate scales is important, because it is the scale at which rearrangements responsible both for thermally activated and for plastic flows occur.  For example, in fragile liquids the boson peak frequency appears to vanish as the glass is heated past its glass transition \cite{tao,chumakov}. This observation suggests that above some temperature an elastic instability occurs in these liquids, and that minima of energy disappear. This scenario was initially proposed by Goldstein \cite{goldstein}, occurs in mean-field models where it strongly affects the dynamics \cite{kurchan}, and has received empirical support in Lennard-Jones \cite{grigera} and colloidal glasses \cite{brito3}. However, which length scales are associated with this instability remains unclear. Mode coupling theory predicts that a dynamical length scale (extractable from a four-point correlation function) should diverge from both sides of this transition \cite{biroli,franz,montanari}, but this is not seen in liquids, where the dynamical length scale continuously grows under cooling. Here  we study the possibility that a  length scale associated with linear elasticity diverges in the solid phase, as the instability is approached.

Elasticity in amorphous materials can be investigated numerically.  Barrat, Tanguy and coworkers have focused in particular on silica, where they showed that a length scale can be consistently extracted from several observables:  the response to a point force, the correlation of non-affine displacements, or the spatial fluctuation of elastic moduli \cite{barrat_lengthscale1,barrat_lengthscale2}. However, questions have remained, of what controls this length scale, and whether or not it is already present in the static structure of the system. Packings of repulsive particles  are  convenient to study this question, because length scales characterizing elasticity become large and even diverge at the unjamming transition where  the pressure vanishes \cite{ohern03,Wyart053,revue,van2010jamming}. In these systems both the  mean number of interactions between the particles (referred to in the following as the coordination $z$) and the applied pressure $p$ play a key role \cite{Wyart052}. The effect of coordination alone can be studied in zero-pressure elastic networks of varying $z$ \cite{wyartmaha,during12,ellenbroek}. %, which share many similarities with sphere packings near the unjamming transition, when $z$ is just above the critical coordination  $z_c = 2d$ predicted by Maxwell \cite{maxwell} ( $d$ denotes the spatial dimension). 
Two length scales appear in such networks at zero pressure. %, but not in the static structure of the material.
 A point-to-set length scale $\ell^*\sim 1/(z-z_c) \equiv 1/\delta z$  characterizes the distance below which mechanical stability of the bulk material is affected by the boundaries \cite{Wyart05}, as observed numerically \cite{during12,bulbul,goodrich2013,tighe12}. Here $z_c = 2d$ is the critical coordination required for stability, as predicted by Maxwell \cite{maxwell}, and $d$ denotes the spatial dimension. Another length scale can be defined by considering the response of the system at the boson peak frequency, numerically one observes a length $\ell_c \sim \delta z^{-1/2}$ \cite{Silbert05,vincenzo10,berthier_jamming} as explained using effective medium \cite{wyart2010}. For floppy networks with $z<z_c$ the response to a zero-frequency force dipole was computed explicitly  \cite{during12}, and was shown to decay on the length $\ell_c$. The same lengthscale characterizes the correlation of non-affine displacements under an imposed global shear \cite{during12}.  These results supported that $\ell^*$ is a point to set length, whereas the response to a local perturbation, as well as two point correlation functions characterizing the response to a global strain, are both characterized by $\ell_c$. However, this interpretation contradicts  early numerical findings supporting that $\ell^*$ characterizes the zero-frequency point response in packings of particles \cite{respprl}, raising doubts on the validity of these numerical results. Moreover, the role of pressure on both $\ell^*$ and $\ell_c$ is currently unclear. 

In this manuscript, we systematically study numerically the response to a local dipolar force in harmonic spring networks and in packings of harmonic soft discs. In agreement with effective medium \cite{wyart2010,during12}, we find that the lengthscale beyond which a continuum elastic description captures correctly the zero-frequency response is actually $\ell_c$, and not $\ell^*$. We demonstrate how increasing pressure in  packings of discs increases  $\ell_c$. Finally, we show data suggesting that $\ell_c$ diverges as the pressure is increased towards the critical pressure at which an elastic  instability occurs. The scaling we find is consistent with $\ell_c(p)\sim 1/(p_c-p)^{1/4}\sim 1/\omega_{BP}(p)$ where $\omega_{BP}(p)$ is the effective medium prediction for the pressure-dependent boson peak frequency as the critical pressure is approached, as derived in a companion paper \cite{unpublished}.

\section{Theoretical framework}
\begin{figure}[!ht]
\centering
\includegraphics[width=0.4\textwidth]{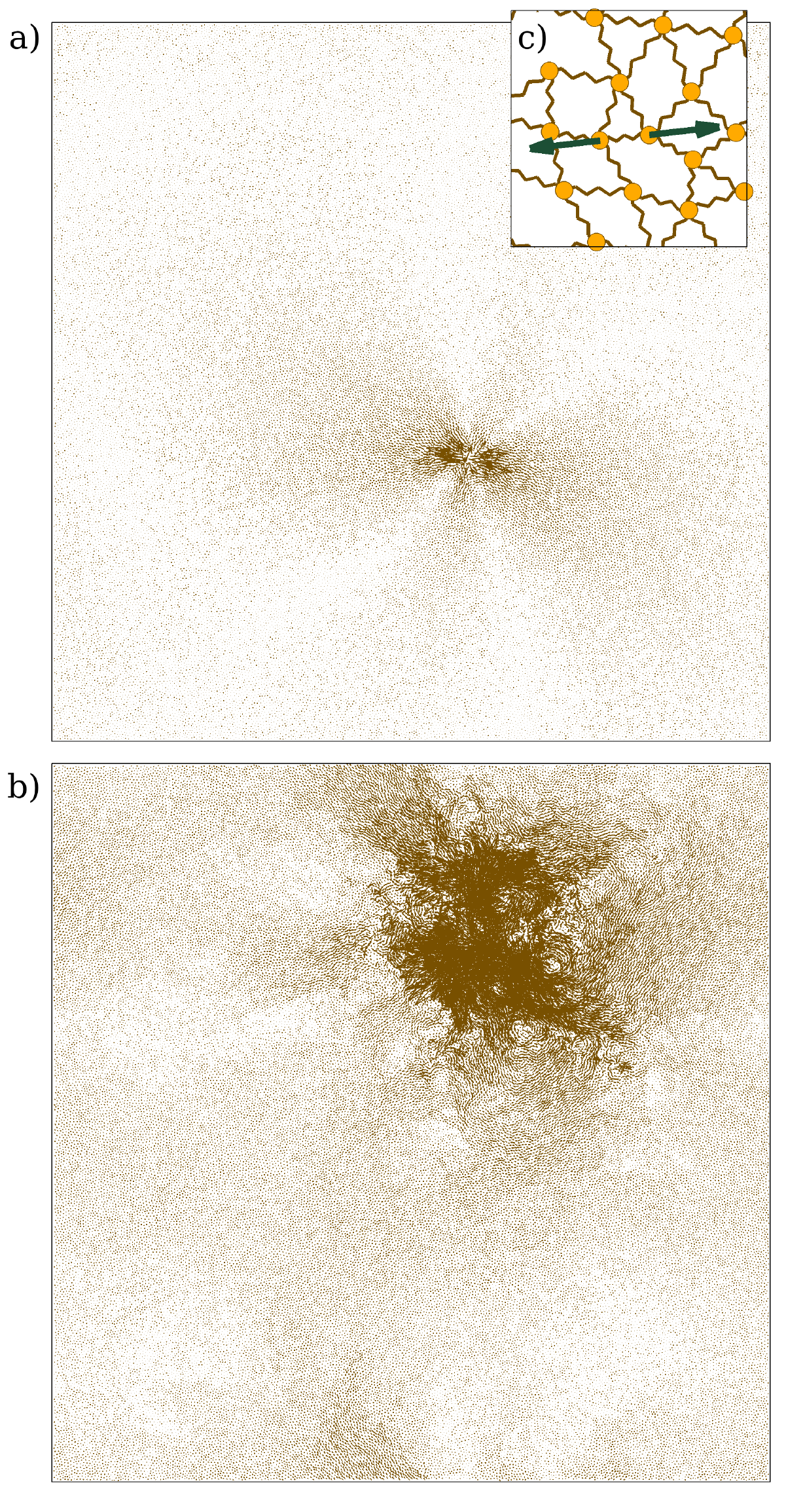}
\caption{Displacement response to a dipolar force, as shown in {(\bf c)}, for spring networks with $N=62500$ nodes, and coordinations $\delta z = 0.8$ ({\bf a}) and $\delta z = 0.05$ ({\bf b}). In this work we extract the lengthscale $\ell_c$ that characterizes this response, and study its dependence on coordination and pressure.}
\label{dipole}
\end{figure}

In this section we provide a general framework within which two response functions to a local dipolar force (as depicted in Fig.~\ref{dipole}c) are defined. The first response function, $C(r)$, measures the amplitude of the change of contact forces  at a distance $r$ away from the perturbation. The second response function, $V(r)$, measures the amplitude of the displacements as a function of $r$. A formal definition of these quantities is presented in this section. The numerical results are presented in Sect.~\ref{results}.

We consider assemblies of $N$ particles interacting via finite-range harmonic pair potentials, with a mean number of interactions per particle $2N_b/N = z > 2d$, with $N_b$ the total number of interactions. We denote by $U$ the potential energy, $\vec{R}_k$ is the position of the $k^{\rm th}$ particle, and the dynamical matrix is $\stackrel{\leftrightarrow}{M}_{jk} \equiv \frac{\partial ^2U}{\partial\vec{R}_j\partial \vec{R}_k}$. We refer to \emph{pairs} of interacting particles as \emph{bonds}.

%%%%%%%%%%%%%%%%%%%%%%%%%%%%%%%%%%%%%%%%%%%%%%%%%%%%%%%
%%%%%%%%%%%%%%%%%%%%%%%%%%%%%%%%%%%%%%%%%%%%%%%%%%%%%%

We next consider a displacement field $\delta \vec{R}_k$ on the coordinates; to linear order in $\delta \vec{R}_k$, this displacement field induces a change $\delta r_{ij}$ in the pairwise distances $r_{ij} \equiv || \vec{R}_j - \vec{R}_i||$ as
\begin{equation}\label{foo2}
\delta r_{ij} \simeq \vec{n}_{ij} \cdot \left( \delta \vec{R}_j - \delta \vec{R}_i\right)\,,
\end{equation}
where $\vec{n}_{ij}$ is the unit vector pointing from $\vec{R}_i$ to $\vec{R}_j$. Eq.(\ref{foo2}) defines a linear operator that takes vectors from the space of the particles' coordinates, to vectors in the space of \emph{bonds}, defined here as the set of pairs of particles that interact. We denote the linear operator defined by Eq.(\ref{foo2}) as ${\cal S}$, and re-write Eq.~(\ref{foo2}) using a bra-ket notation:
\begin{equation}
\ket{\delta r} \simeq {\cal S}\ket{\delta R}\,.
\end{equation}
We next consider a set of forces $f_{ij}$ on each bond $\langle ij \rangle$, and compute the net force $\vec{F}_k$ that results from the bond-forces exerted on the $k^{\rm th}$ particle as
\begin{equation}\label{foo3}
\vec{F}_k = \sum_{j(k)}\vec{n}_{jk}f_{jk}\,,
\end{equation}
where $j(k)$ denotes the set of all particles $j$ that interact with particle $k$. Similarly to Eq.(\ref{foo2}), Eq.(\ref{foo3}) also defines a linear operator, but this time it takes vectors from the space of bonds, to vectors in the space of particles' coordinates. It is easy to show \cite{simulation_paper} that it is, in fact, the transpose of the operator ${\cal S}$ which is defined by Eq.(\ref{foo3}), and we can therefore write Eq.(\ref{foo3}) in bra-ket notation as
\begin{equation}
\ket{F} = {\cal S}^T\ket{f}\,.
\end{equation}

% We next define the linear operator ${\cal S}$ and its transpose ${\cal S}^T$ via their operations on general vectors $\ket{V}$ and $\ket{f}$:
% \begin{eqnarray}
% \vec{n}_{ij}\cdot(\vec{V}_j - \vec{V}_i) \ \ &\longleftrightarrow &\ \  {\cal S}\ket{V}\ , \label{es_definition} \\
% \sum_{i(j)}\vec{n}_{ij}f_{ij} \ \ &\longleftrightarrow &\ \  {\cal S}^T\ket{f}\ , \label{es_transpose_definition} 
% \end{eqnarray}
% where $\vec{n}_{ij}$ are the unit vectors defined for each bond $\langle ij \rangle$, and point from the center of particle $i$ to the center of particle $j$, and $i(j)$ denotes the set of all the particles $i$ that are interacting with particle $j$. Notice that the vectors $\ket{V}$ and ${\cal S}^T\ket{f}$ are of dimension $N\times d$ (the dimension of the space of coordinates), but the vectors $\ket{f}$ and ${\cal S}\ket{V}$ are of dimension $N_b$ (the dimension of the space of bonds). 

%%%%%%%%%%%%%%%%%%%%%%%%%%%%%%%%%%%%%%%%%%%%%%%%%%%%%%%
\begin{figure*}[ht]
%\begin{center}
\includegraphics[width=1.0\textwidth]{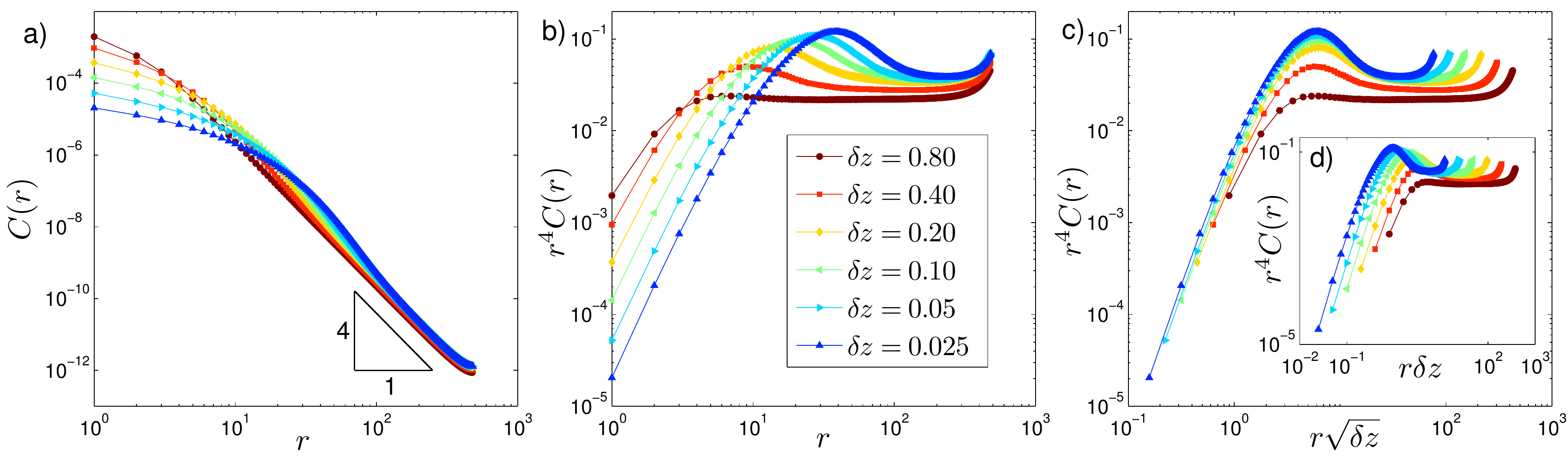}
\caption{{\bf a)}~Response functions $C(r)$ for spring networks in two dimensions at coordinations as indicated by the legend. {\bf b)}~Plotting $r^4C(r)$ reveals the continuum linear elastic behvior at large $r$, and indicates that as $\delta z \to 0$, $C(r) \sim c/r^{-2d}$ with $c$ independent of $z$. {\bf c)} Plotting $r^4C(r)$ \emph{vs.}~$r\sqrt{\delta z}$ results in the alignment of the peaks of $r^4C(r)$ which indicates that the lengthscale dominating this response is $\ell_c\sim 1/\sqrt{\delta z}$. {\bf d)} Plotting $r^4C(r)$ \emph{vs.}~the rescaled length $r/\ell^*\sim r\delta z$ does not lead to the alignment of the peaks, reinforcing that $\ell_c \sim 1/\sqrt{\delta z}$ is the relevant lengthscale.}
\label{noPrestress}
%\end{center}
\end{figure*}
%%%%%%%%%%%%%%%%%%%%%%%%%%%%%%%%%%%%%%%%%%%%%%%%%%%%%%%

We next define $\ket{\alpha}$ as a vector in the space of bonds that has zeros in all components, and has one for the $\alpha$ component, which corresponds to a single bond $\langle ij \rangle$. The operation of ${\cal S}^T$ on the vector $\ket{\alpha}$ is a coordinate-space dipole vector that can be expressed as $(\delta_{jk}-\delta_{ik})\vec{n}_{ij}$, whose squared magnitude is 
\begin{equation}
\bra{\alpha}{\cal S}{\cal S}^T\ket{\alpha} = \sum_k (\delta_{jk} - \delta_{ik})(\delta_{jk} - \delta_{ik})\vec{n}_{ij}\cdot\vec{n}_{ij}=2\,.
\end{equation}
An example of the vector ${\cal S}^T\ket{\alpha}$ is depicted in Fig.~\ref{dipole},c. We consider now the displacement response $\ket{\delta R^{(\alpha)}}$ to a dipolar force ${\cal S}^T\ket{\alpha}$:
\begin{equation}\label{displacementResponse}
\ket{\delta R^{(\alpha)}} = {\cal M}^{-1}{\cal S}^T\ket{\alpha}\,,
\end{equation}
with ${\cal M}$ the dynamical matrix. Two examples of $\ket{\delta R^{(\alpha)}}$ for spring networks (see details below) are shown in Fig.~(\ref{dipole},a,b). In this work we consider two response functions extracted from these displacement responses to dipolar forces in simple amorphous solids.

Before defining the first response function, we note that the displacement response $\ket{\delta R^{(\alpha)}}$ changes the distance between particles in the entire system. In particular, the change in distance between the particles that form the bond $\beta$, to first order in $||\delta R^{(\alpha)}||$, is given by
\begin{equation}\label{define_A}
\bra{\delta R^\alpha}{\cal S}^T\ket{\beta} = \bra{\alpha}{\cal S}{\cal M}^{-1}{\cal S}^T\ket{\beta} \equiv \bra{\alpha}{\cal A}\ket{\beta}\, .
\end{equation}
Eq.~(\ref{define_A}) defines a symmetric, non-negative definite linear operator ${\cal A}$ of dimension $N_b \times N_b$,
which operates on vectors in the space of bonds. We leave the eigenmode analysis of this operator for future work. The matrix elements $\bra{\alpha}{\cal A}\ket{\beta}$ depend on the distance $r$ between the bonds $\alpha$ and $\beta$, which can be defined as the distance between the mean position of the particles that form each bond. We now define the response function
\begin{equation}\label{responseFunction}
C(r) \equiv \left[ \bra{\alpha}{\cal A}\ket{\beta}^2 \right]_r\, ,
\end{equation}
where the square brackets denoting averaging over all pairs of bonds $\alpha,\beta$ that are separated by a distance $r$. Continuum linear elasticity predicts $C(r) \sim r^{-2d}$, since ${\cal A}$ scales as the gradient of the displacement response, and the latter decays away from the perturbation as $r^{1-d}$. 

The second response function we consider in the following is the square of the displacement response at a distance $r$ away from the dipolar force, namely
\begin{equation}
V(r) \equiv \left[  ||\delta\vec{R}^{(\alpha)}_k||^2 \right]_r \,,
\end{equation}
where this time the square brackets denotes averaging over all particles $k$ that are located at a distance $r$ from the dipolar force applied to the bond $\alpha$. Continuum linear elasticity predicts $V(r) \sim r^{2(1-d)}$, as it is the square of the displacement response, which, as noted above, decays away from the perturbation as $r^{1-d}$. 

In the following we use the continuum linear elastic predictions $C(r) \sim r^{-2d}$ and $V(r)\sim r^{2(1-d)}$ to extract the lengthscale $\ell_c$.

\section{Results}
\label{results}
In this work we focus on assemblies of particles interacting via harmonic pair-potentials: spring networks and disc packings. The energy is $U = \sum_{i<j}\phi_{ij}$, $\phi_{ij} = \tilde{k}(r_{ij} - d_{ij})^2$ with $\tilde{k}$ a spring constant (set in the following to unity), $d_{ij} = (d_i + d_j)/2$ for harmonic discs with diameters $d_k$, and $d_{ij}$ is the restlength of the $\langle ij \rangle$ spring for the spring networks.

\subsection{Spring networks}
We consider first spring networks in which all of the springs are at their respective rest-lengths, which implies that there are no internal stresses in the system. Networks of up to $N=10^6$ nodes in two dimensions were prepared following the protocol described in \cite{wyartmaha}, which results in networks having small fluctuations of the coordination amongst the nodes. The mean spring length defines our unit of length.

The response functions $C(r)$ are presented in Fig.~\ref{noPrestress} for networks of $N=10^6$ nodes and various coordinations as indicated by the legend. In panel {\bf a)} we plot the raw functions $C(r)$, which indeed seem to obey the asymptotic linear-elastic prediction $C(r) \sim r^{-2d}$. The prefactor of this scaling seems to converge to a constant as $\delta z\to 0$, as can be seen in panel {\bf b)}, where the products $r^{2d}C(r)$ are plotted. The increase at large $r$ is an effect of the periodic boundary conditions. In Fig.~\ref{noPrestress}c we plot the products $r^{2d}C(r)$ \emph{vs}.~the rescaled length $r\sqrt{\delta z}$. The alignment of the peaks of the response functions validates that the lengthscale dominating the response to a local dipolar force is $\ell_c \sim 1/\sqrt{\delta z}$. Beyond $l_c$ we find a plateau as expected for a continuous elastic medium. The lack of alignment when $r^{2d}C(r)$ is plotted against the rescaled length $r\delta z$ (see panel {\bf d)}) supports that $\ell_c$ is the relevant lengthscale for this response, and not $\ell^*\sim1/\delta z$. 

%%%%%%%%%%%%%%%%%%%%%%%%%%%%%%%%%%%%%%%%%%%%%%%%%%%%%%%
\begin{figure}[!ht]
\centering
\includegraphics[scale = 0.53]{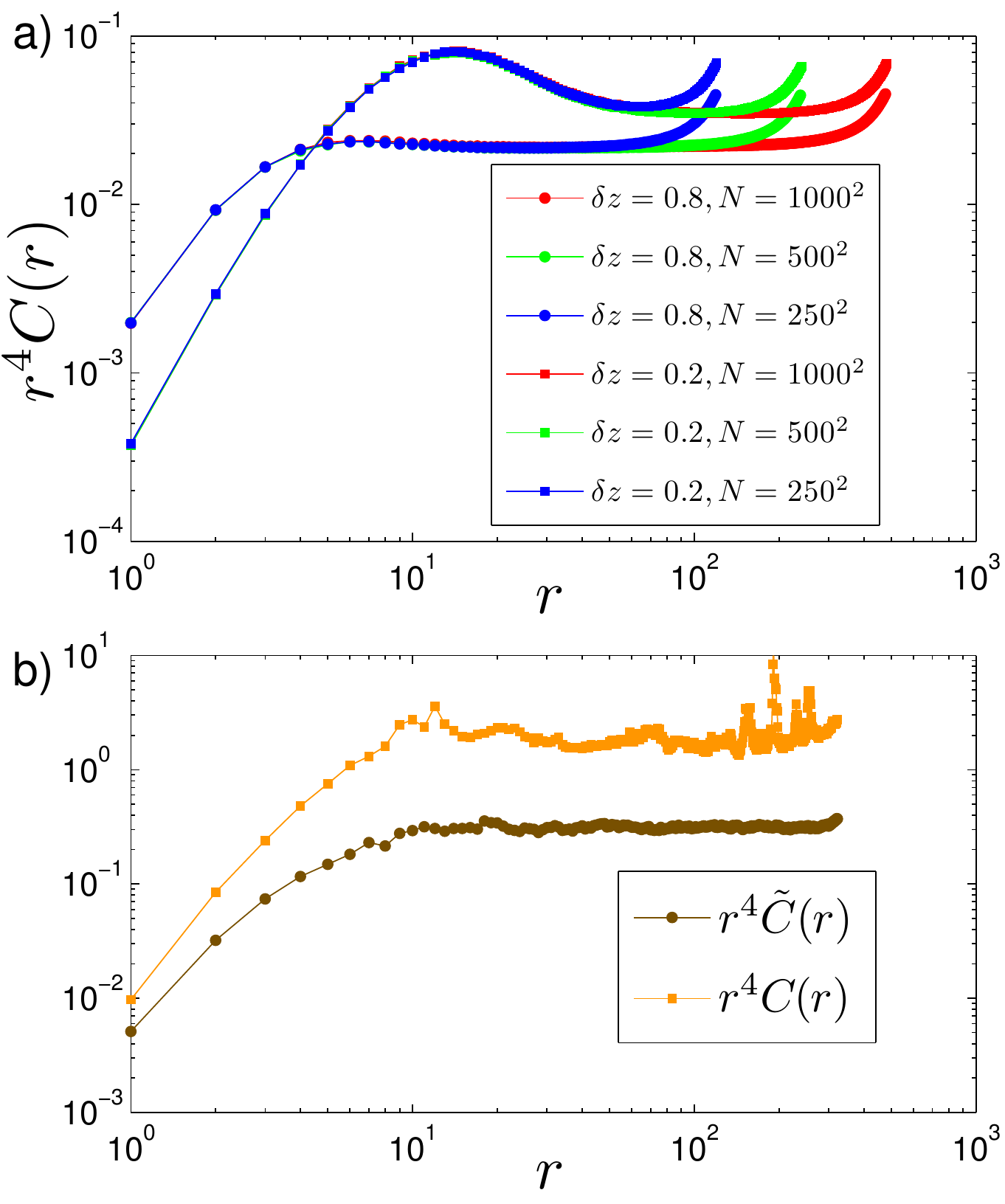}
\caption{{\bf a)}~The products $r^4C(r)$ measured in two-dimensional spring networks of various sizes and coordinations as indicated by the legend. {\bf b)}~The products $r^4C(r)$ (squares) and $r^4\tilde{C}(r)$ (circles) measured in packings of harmonic discs at the pressure $p=10^{-1}$. Both functions plateau at the same lengthscale; however, it is apparent that $\tilde{C}(r)$ smooths out the noise seen in $C(r)$.}
\label{system_size_independence}
\end{figure}
%%%%%%%%%%%%%%%%%%%%%%%%%%%%%%%%%%%%%%%%%%%%%%%%%%%%%%

In Fig.~\ref{system_size_independence}a we plot the products $r^4C(r)$ for coordinations $\delta z = 0.8$ and $\delta z = 0.2$, measured in networks of $N=250^2$, $N=500^2$ and $N=1000^2$ nodes in two dimensions. We find that the system size has no effect on the response functions at lengths smaller than half the linear size of the system $r<L/2$. This result demonstrates the validity of our procedure to extract the lengthscale $\ell_c$ from the positions of the peaks of the response functions. 

%%%%%%%%%%%%%%%%%%%%%%%%%%%%%%%%%%%%%%%%%%%%%%%%%%%%%%%
\begin{figure}[!ht]
\centering
\includegraphics[scale = 0.62]{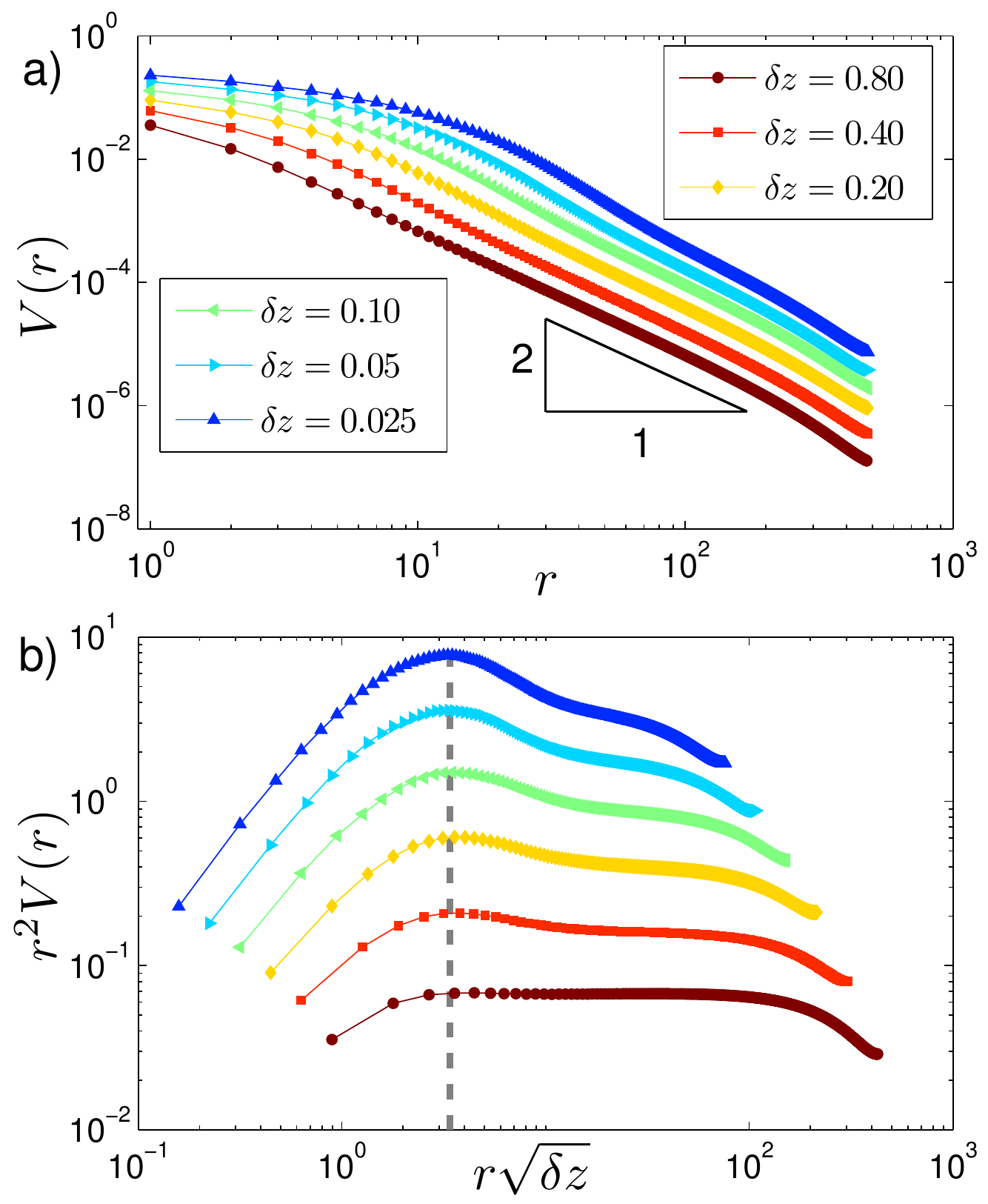}
\caption{{\bf a)}~The response functions $V(r)$ for spring networks in two dimensions at coordinations as indicated by the legend. {\bf b)}~The products $r^2V(r)$ are plotted \emph{vs}.~the rescaled length $r\sqrt{\delta z}$. The vertical dashed line demonstrates that the peaks of the response functions for different coordinations align when plotted against the rescaled length.}
\label{displacementsFig}
\end{figure}
%%%%%%%%%%%%%%%%%%%%%%%%%%%%%%%%%%%%%%%%%%%%%%%%%%%%%%

In Fig.~\ref{displacementsFig} we plot the response functions $V(r)$ measured in spring networks. Panel {\bf a)} of Fig.~\ref{displacementsFig} displays the raw functions, which appear to obey the continuum linear elastic prediction $V(r)\sim r^{2(1-d)}$ at sufficiently large $r$. In panel {\bf b)} of Fig.~\ref{displacementsFig} the products $r^2V(r)$ are plotted vs.~the rescaled length $r\sqrt{\delta z}$. The alignment of the peaks demonstrates that the lengthscale characterizing this response function is also $\ell_c \sim 1/\sqrt{\delta z}$. 

To rationalize these findings, we estimate the sum of squares of the displacement response to a local dipolar force (see Eq.~(\ref{displacementResponse})), $||\delta R^{(\alpha)}||^2$, as
\begin{equation}\label{v2}
|| \delta R^{(\alpha)}||^2 = \braket{\delta R^{(\alpha)}}{\delta R^{(\alpha)}} = \bra{\alpha}{\cal S}{\cal M}^{-2}{\cal S}^T\ket{\alpha}\,.
\end{equation}
We denote by $\ket{\Psi_\omega}$ the eigenmode of ${\cal M}$ with the associated eigenvalue $\omega^2$. For our unstressed elastic networks with $\tilde{k}=1$, ${\cal M} = {\cal S}^T{\cal S}$ \cite{Wyart053}, and we define $\omega\ket{\psi_\omega}\equiv{\cal S}\ket{\Psi_\omega}$. The bond-space vectors $\ket{\psi_\omega}$ are normalized:
\begin{equation}
\braket{\psi_\omega}{\psi_{\omega}} = \frac{\bra{\Psi_\omega}{\cal S}^T{\cal S}\ket{\Psi_\omega}}{\omega^2} = 
\frac{\bra{\Psi_\omega}{\cal M}\ket{\Psi_\omega}}{\omega^2} = 1\ .
\end{equation}
We can thus write Eq.~(\ref{v2}) as 
\begin{equation}
|| \delta R^{(\alpha)}||^2  = \sum_\omega\frac{\bra{\alpha}{\cal S}\ket{\Psi_\omega}^2}{\omega^4} = \sum_\omega\frac{\braket{\alpha}{\psi_\omega}^2}{\omega^2}\,.
\end{equation}
The normalization of the vectors $\ket{\psi_\omega}$ implies that, upon averaging over contacts $\alpha$, $\braket{\alpha}{\psi_\omega}^2\sim N^{-1}$, and we can approximate the sum over eigenfrequencies by an integral over the density of states $D(\omega)$:
\begin{equation}\label{foo1}
|| \delta R^{(\alpha)}||^2 \approx \int \frac{D(\omega)d\omega}{\omega^2}\,.
\end{equation}
In our unstressed elastic networks theory predicts \cite{Wyart05,wyart2010} $D(\omega)\sim \delta z^{-d/2}\omega^{d-1}$ for $\omega < \omega^* \sim \delta z$, and $D(\omega) \sim \mbox{constant}$ for $\omega \gtrsim \omega^*$. The lowest mode is expected to be a plane-wave with a frequency of order $\sqrt{\mu}/L$ with shear modulus $\mu \sim \delta z$ , and $L$ the linear size of the system. We decompose the integral of Eq.~(\ref{foo1}) as
\begin{eqnarray}
|| \delta R^{(\alpha)}||^2 & \approx & \int \frac{D(\omega)d\omega}{\omega^2}  \\ \nonumber
& \sim & \delta z^{-d/2}\int\limits_{\sqrt{\mu}/L}^{\omega^*}\omega^{d-3}d\omega + \int\limits_{\omega*}^1\omega^{-2}d\omega \\ \nonumber
& \sim & \left\{ \begin{array}{cc} \frac{1}{\delta z} &,\  d\ge3 \\ \frac{1}{\delta z}\left(1+B\log(L/\ell_c)\right) &,\  d=2  \end{array}\right. \,.
\end{eqnarray}
We find that $|| \delta R^{(\alpha)}||^2$ is dominated by the modes at~$\omega^*$ (in $d\ge3$, with logarithmic corrections in $d=2$), which are correlated on the correlation length $\ell_c$ according to effective medium \cite{wyart2010,unpublished}. We therefore expect observables derived from the response to a local dipolar force, such as ~$V(r)$ or $C(r)$, to be characterized by that scale, as we indeed find.

% $\ket{\psi_\omega}$  is defined as the operation of ${\cal S}$ on eigenmodes $\ket{\Psi_\omega}$. For our unstressed elastic networks with $\tilde{k}=1$, it is easy to show \cite{} that
% \begin{eqnarray}
% {\cal M} & = & {\cal S}^T{\cal S} = \sum_\omega\omega^2\ket{\Psi_\omega}\bra{\Psi_\omega}\,, \\ 
% {\cal M} & = & {\cal S}^T{\cal S} = \sum_\omega\omega^2\ket{\Psi_\omega}\bra{\Psi_\omega}\,.
% \end{eqnarray}

\subsection{Packings of harmonic discs}
We next show results of a similar analysis performed on two dimensional packings of soft discs. Packings of $N=360,000$ bi-disperse harmonic discs with a diameter ratio 1.4 of were created by quenching a high-temperature fluid to zero temperature using the FIRE algorithm \cite{FIRE}, and applying compressive or expansive strains followed by additional quenches to obtain the target pressures. The diameter of the small discs $d_0$ are taken as our units of length, so that forces and pressure are measured in units of $\tilde{k}d_0$ and $\tilde{k}d_0^{2-d}$ respectively.

Unlike unstressed spring networks, in packings particles exert contact forces on each other. These forces are known to destabilize packings \cite{Wyart052}, and indeed they give rise to much noisier responses compared to the unstressed networks. To deal effectively with this noise, we define the response function $\tilde{C}(r)$:
\begin{equation}
\tilde{C}(r) \equiv {\rm median}_r\left( \bra{\alpha}{\cal A}\ket{\beta}^2\right)\,,
\end{equation}
where the median is taken over all pairs of contacts $\alpha,\beta$ that are separated by a distance $r$ from each other. In Fig.~\ref{system_size_independence}b both response functions $C(r)$ and $\tilde{C}(r)$ measured in packings at the pressure $p=10^{-1}$ are compared.

%%%%%%%%%%%%%%%%%%%%%%%%%%%%%%%%%%%%%%%%%%%%%%%%%%%%%%%
%\begin{figure}[!ht]
\begin{figure}
\centering
\includegraphics[scale = 0.6]{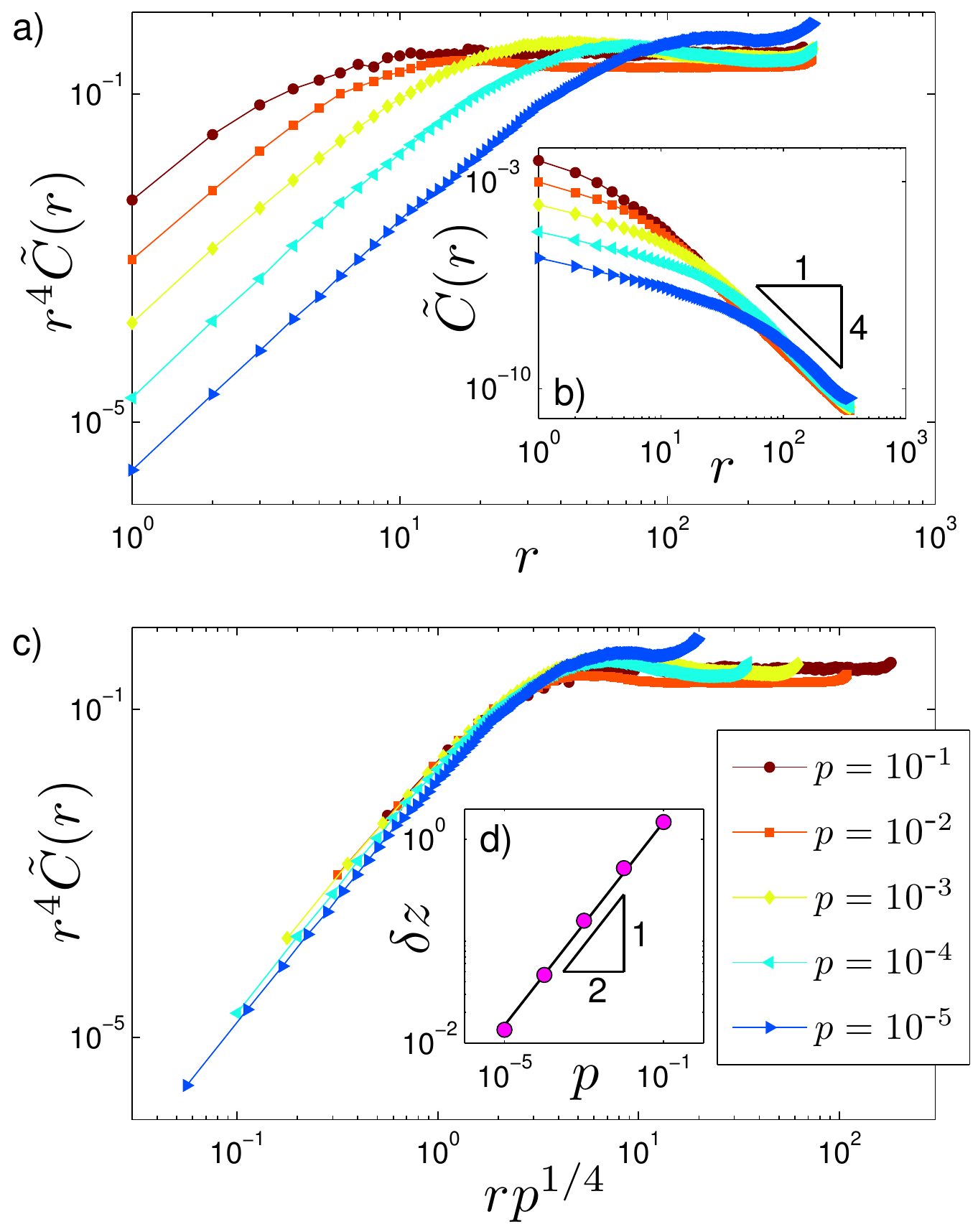}
\caption{{\bf a)} The products $r^4\tilde{C}(r)$ measured in packings of $N=360,000$ harmonic discs in two dimensions. {\bf b)} Raw response functions for packings, which scale as $\tilde{C}(r) \sim r^{-2d}$ at large $r$. The products $r^{2d}\tilde{C}(r)$ plotted against the rescaled variable $rp^{-1/4}$ reveal the lengthscale $\ell_c\sim\delta z^{-1/2}$.}
\label{packings_length}
\end{figure}
%%%%%%%%%%%%%%%%%%%%%%%%%%%%%%%%%%%%%%%%%%%%%%%%%%%%%%
%%%

In Fig.~\ref{packings_length} we plot the response functions $\tilde{C}(r)$ measured in our two-dimensional packings at pressures indicated by the legend. We find that although the shape of the response function $\tilde{C}(r)$ slightly differs from $C(r)$ measured in spring networks, the main features are similar, and in particular a crossover to $\tilde{C}(r) \sim r^{-2d}$ occurs on the scale $\ell_c \sim p^{-1/4}\sim \delta z^{-1/2}$ (the coordination in harmonic packings scales as $\delta z\sim\sqrt{p}$ \cite{ohern03, Wyart052}, verified in the data of panel~{\bf d} of Fig.~\ref{packings_length}). Surprisingly, we find that the fluctuations in $\tilde{C}(r)$ are largest for the highest pressure $(p=10^{-1})$; we leave the investigation of the nature of these fluctuations for future work.
%as can also be seen in Fig.~\ref{system_size_independence},b. 

\subsection{Effect of internal stresses}

To directly probe the effect of internal stresses on the response to a local dipolar force, we prepared packings of $N=10^6$ harmonic discs at the packing fraction $\phi = 0.86$, which have a mean coordination of $z \approx 4.4$ and mean pressure of $p_0 \approx 7.6\times10^{-3}$. We then consider the response function $\tilde{C}^{(x)}(r)$ to a local dipolar force, which is calculated with a dynamical matrix in which the contact forces are multiplied by a factor $1-x$, namely
\begin{eqnarray}\label{em_of_x}
\stackrel{\leftrightarrow}{M}^{(x)}_{mq}& = &\sum_{\langle ij \rangle}(\delta_{mj}-\delta_{mi})(\delta_{qj}-\delta_{qi})\left[\tilde{k}\vec{n}_{ij}\vec{n}_{ij} \right. \nonumber \\
&& \left. \quad\quad - (1-x)\left(\stackrel{\leftrightarrow}{I} - \vec{n}_{ij}\vec{n}_{ij}\right)f_{ij}/r_{ij}\right] \, ,
\end{eqnarray}
where $\stackrel{\leftrightarrow}{I}$ is the unit tensor, and the sum is over all contacts ${\langle ij \rangle}$. The original dynamical matrix (and hence the original response function $\tilde{C}(r)$) is recovered for $x=0$. This rescaling of the forces leads to the rescaling of the pressure $p = (1-x)p_0$ where $p_0$ is the pressure of the original packing. 

%%%%%%%%%%%%%%%%%%%%%%%%%%%%%%%%%%%%%%%%%%%%%%%%%%%%%%%
%\begin{figure}[!ht]
\begin{figure}
\centering
\includegraphics[scale = 0.6]{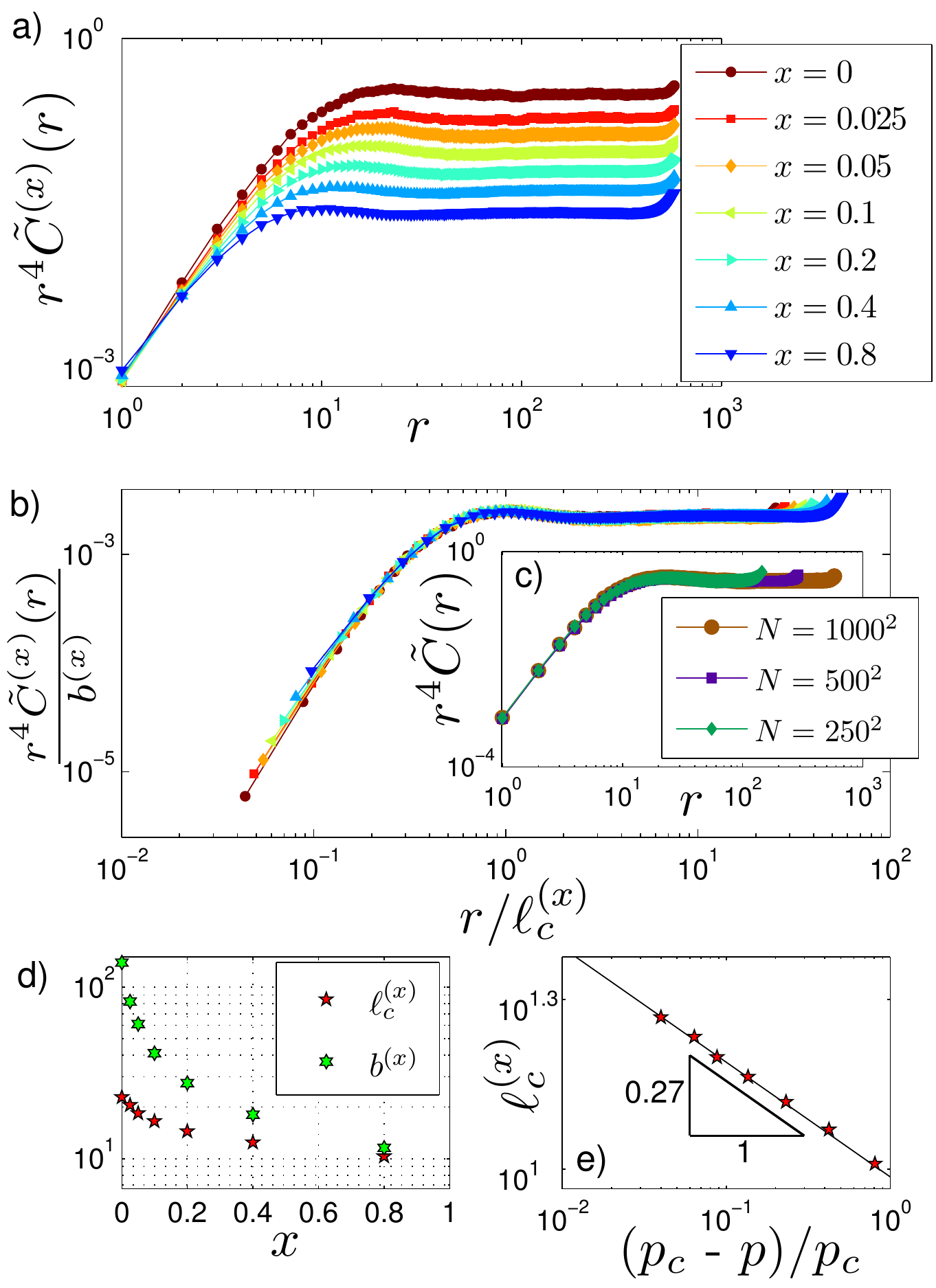}
\caption{{\bf a)}~The products $r^{2d}\tilde{C}^{(x)}(r)$ measured in packings of $N=10^6$ harmonic discs at the packing fraction $\phi = 0.86$, for various factors $x$, see text for details. {\bf b)}~Plotting the~rescaled products $r^{2d}\tilde{C}^{(x)}(r)/b^{(x)}$ \emph{vs.}~$r/\ell_c^{(x)}$ leads to a collapse for all $x$, from which we extract the lengthscales $\ell_c^{(x)}$ and the amplitudes $b^{(x)}$, which are plotted in panel {\bf d)} on a semi-log scale. {\bf e)} The lengthscale $\ell_c^{(x)}$ \emph{vs.}~the relative proximity to~the critical pressure $p_c$, see text for details. {\bf c)}~The products $r^{2d}\tilde{C}(r)$ measured for packings at $\phi=0.86$ and various system sizes.}
\label{prestress_length_fig}
\end{figure}
%%%%%%%%%%%%%%%%%%%%%%%%%%%%%%%%%%%%%%%%%%%%%%%%%%%%%%

The products $r^{2d}\tilde{C}^{(x)}(r)$ are plotted in Fig.~\ref{prestress_length_fig} for various values of $x$ as indicated by the legend. Here we find that the response is governed by an $x$-dependent lengthscale $\ell_c^{(x)}$, which is extracted by rescaling the axes by the appropriate lengths and amplitudes (plotted in panel {\bf d)} of Fig.~\ref{prestress_length_fig}) such that the curves collapse. For distances $r > \ell_c^{(x)}$, $\tilde{C}^{(x)}(r) \sim b^{(x)}r^{-2d}$, with an $x$-dependent prefactor $b^{(x)}$.

Our results indicate clearly that the lengthscale $\ell_c$ that governs the response to a local dipolar force is sensitive to the presence of internal stresses in the solid, and, in particular, it decreases as the pressure is decreased by rescaling the contact forces. However, the question remains whether $\ell_c$ exhibits singular behavior as the internal stresses are increased. We  clarify this issue by considering again the dynamical matrix ${\cal M}^{(x)}$ given by Eq.~(\ref{em_of_x}), and denoting the density of states associated to ${\cal M}^{(x)}$ by $D^{(x)}(\omega)$. In the companion paper~\cite{unpublished} it is shown that $D^{(x)}(\omega)$ displays an $x$-dependent frequency scale $\omega_0(x)$, which characterizes the destabilizing effect of internal stresses. It is shown that for $x_c \approx -0.04$, $\omega_0(x_c)$ vanishes, which corresponds to an elastic instability \cite{Wyart052,unpublished}. $x_c$ defines a critical pressure $p_c = (1-x_c)p_0$ at which this elastic instability occurs. The relative distance of the pressure $p = (1-x)p_0$ from the critical pressure $p_c$, given $x$, is thus
\begin{equation}
\frac{p_c - p}{p_c} = \frac{x-x_c}{1-x_c}\ .
\end{equation}
When we plot the extracted lengthscale $\ell_c^{(x)}$ \emph{vs}.~the relative distance to the critical pressure $(p_c - p)/p_c$, we find $\ell_c^{(x)} \sim \left( \frac{p_c - p}{p_c} \right)^{-0.27}$, using $x_c = -0.04$, see panel {\bf e)} of Fig.~\ref{prestress_length_fig}. This result suggests that $(i)$ our original harmonic disc packings dwell at a pressure $p_0$ that is a fraction $(1-x_c)^{-1}\approx 96\%$ of the critical pressure $p_c$, i.e. very close to marginal stability \cite{Wyart053,Wyart052}, and $(ii)$ the lengthscale $\ell_c^{(x)}$ diverges at the critical pressure $p_c$. Although our variation of $\ell_c^{(x)}$ is mild due to the smallness of the exponent, there is no fitting involved once $x_c$ is independently determined,  supporting that the power-law  is genuine. 

We finally discuss the possibility that the critical pressure $p_c$ approaches the pressure $p_0$ at which our packings dwell in the thermodynamic limit. If indeed $\ell_c^{(x)}$ diverges with proximity to the critical pressure $p_c$, assuming that $p_c\to p_0$ as $N\to\infty$ would imply that $\ell_c^{(x)}$ should increase as $N$ is increased. To rule out this possibility, we plot in panel {\bf c)} of Fig.~\ref{prestress_length_fig} the products $r^4\tilde{C}(r)$ measured in packings of $N=250^2, N=500^2$, and $N=1000^2$, at the packing fraction $\phi = 0.86$. We find that the lengthscale $\ell_c = \ell_c^{(x=0)}$ which characterizes these response functions does not change with increasing the system size $N$. 

\section{discussion}

In elastic networks at zero pressure, our results support our earlier claim \cite{during12} that  $\ell^*\sim 1/\delta z$  is  a point-to-set length associated with the dependence of mechanical stability on pinning or cutting boundaries, whereas the length $\ell_c\sim1/\sqrt{\delta z}$ characterizes the zero-frequency point response and two-point correlation functions, as justified here based on former results from effective medium. Note that the presence of two length scales is not related to the fact that in particle packings the longitudinal and transverse speed of sound scale differently, as proposed in \cite{goodrich2013}. Indeed in elastic networks both the shear and bulk moduli can be made to scale identically \cite{Wyart053,ellenbroek}, and these two length scales still differ. More work is needed to understand which of these length scales characterize certain finite size effects \cite{goodrich2013,moukarzel2,during12}.

When pressure is increased toward a critical value $p_c$ at which the system becomes elastically unstable, we expect both $\ell^*$ and $\ell_c$ to grow. Our results are consistent with a divergence of  $\ell_c\sim 1/(p_c-p)^{1/4}$. Our results (see also \cite{unpublished}) suggest that sphere packings are very close to, but at a finite distance from, an elastic instability with $(p_c-p)/p_c\approx 0.05\%$ independently of coordination, implying that $\ell_c$ in elastic networks at rest and in packings are proportional, thus depending on coordination as $\ell_c\sim 1/\sqrt{\delta z}$ in both cases. %The dependence of $\ell^*$ with pressure could be tested numerically by opening boundaries at some distance $L$ while maintaining fixed particle positions. $\ell^*(p)$ would then be the value of $L$ below which unstable modes penetrate in the center of the material. 
Our prediction of a diverging length scale near an elastic instability could be tested in various contexts, for example near an amorphization transition where the distance to the instability can be controlled by monitoring disorder \cite{mizuno2013elastic}, and experimentally in shaken grains \cite{coulais} or colloidal systems \cite{chen2010low}. 

These results resemble predictions of Mode Coupling Theory (MCT), believed to describe some kind of elastic instability \cite{biroli,franz,montanari}. MCT predicts a {\it dynamical} length scale $\xi$ diverging as $\xi\sim |T-T_c|^{-1/4}$. This is the same exponent as in our observation of a  length $\ell_c\sim (p_c-p)^{-1/4}$ characterizing the zero-frequency response, since pressure and temperature should be linearly related. We observe that this scaling of  $\ell_c(p)$ goes as the inverse boson peak frequency $\omega_{BP}(p)$ as predicted using effective medium in a companion paper \cite{unpublished}, but this correspondence is currently unexplained. Overall, a systematic comparison between mode coupling and effective medium -in particular on the length scales involved in each approach- would be  valuable to clarify the relationship between dynamics and elasticity in supercooled liquids.

\acknowledgements
 Acknowledgments: We thank Jie Lin, Le Yan and Marija Vucelja for discussions. MW acknowledges support from NSF CBET Grant 1236378, NSF DMR Grant 1105387, and MRSEC Program of the NSF DMR-0820341 for partial funding. GD~acknowledges support from CONICYT  PAI/Apoyo al Retorno 82130057.

\bibliography{reference10}{}
\end{document}